\pdfoutput=1

\documentclass[11pt]{article}

\usepackage{ACL2023}
\usepackage{times}
\usepackage{latexsym}

\usepackage[T1]{fontenc}

\usepackage[utf8]{inputenc}

\usepackage{microtype}

\usepackage{inconsolata}
\usepackage{graphicx}
\usepackage{float} 
\usepackage{subfig}
\usepackage{amsmath}
\title{RMSSinger: Realistic-Music-Score based Singing Voice Synthesis}

\author{Jinzheng He$^{1}$, Jinglin Liu$^{2}$,  Zhenhui Ye$^{1}$, Rongjie Huang$^{1}$, Chenye Cui$^{1}$, Huadai Liu$^{1}$, Zhou Zhao$^{1}$
\thanks{~~Corresponding author.}\\
$^{1}$Zhejiang University \\
\texttt{\{jinzhenghe,zhenhuiye,rongjiehuang,chenyecui,huadailiu,zhaozhou\}@zju.edu.cn} \\
$^{2}$ByteDance\\
\texttt{liu.jinglin@bytedance.com}\\
} 

\begin{document}
\maketitle
\begin{abstract}
We are interested in a challenging task, \textbf{R}ealistic-\textbf{M}usic-\textbf{S}core based \textbf{S}inging \textbf{V}oice \textbf{S}ynthesis (RMS-SVS). RMS-SVS aims to generate high-quality singing voices given realistic music scores with different note types (grace, slur, rest, etc.). Though significant progress has been achieved, recent singing voice synthesis (SVS) methods are limited to fine-grained music scores, which require a complicated data collection pipeline with time-consuming manual annotation to align music notes with phonemes. 
Furthermore, these manual annotation destroys the regularity of note durations in music scores, making fine-grained music scores inconvenient for composing.
To tackle these challenges, we propose RMSSinger, the first RMS-SVS method, which takes realistic music scores as input, eliminating most of the tedious manual annotation and avoiding the aforementioned inconvenience.
Note that music scores are based on words rather than phonemes, in RMSSinger, we introduce word-level modeling to avoid the time-consuming phoneme duration annotation and the complicated phoneme-level mel-note alignment. Furthermore, we propose the first diffusion-based pitch modeling method, which ameliorates the naturalness of existing pitch-modeling methods. To achieve these, we collect a new dataset containing realistic music scores and singing voices according to these realistic music scores from professional singers. Extensive experiments on the dataset demonstrate the effectiveness of our methods. Audio samples are available at \url{https://rmssinger.github.io/}.
\end{abstract}

\begin{figure}[!h]
\centering
\subfloat[][fine-grained music scores]{\includegraphics[width=0.9\linewidth]{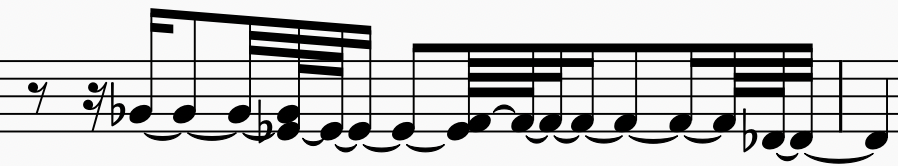}}

\subfloat[][realistic music scores]{\includegraphics[width=0.9\linewidth]{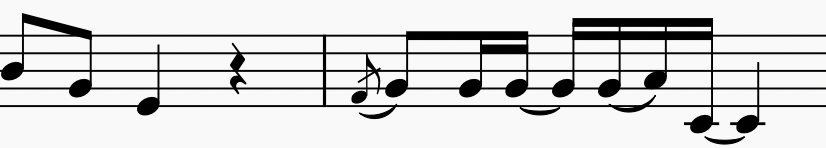}}
\caption{Difference between fine-grained and realistic music scores. Manual adjustment destroys the regularity of note durations, making fine-grained music scores "crushed" and inconvenient for composing.}
\label{fig_data}
\vspace{-0.3cm}
\end{figure}

\section{Introduction}
Singing Voice Synthesis (SVS) aims to generate high-quality singing given music scores (lyrics, note pitches, and note durations), and has attracted increasing academic and industrial attention. SVS is extensively required in both professional music composing and entertainment industries in real life\citep{umbert2015expression}. 

Though significant progress has been achieved, recent SVS methods\citep{wang2022opencpop,zhang2022wesinger,liu2022diffsinger,zhangm4singer,huang2021multi,huang2022singgan} cannot utilize realistic music scores from composers or websites but require fine-grained music scores. Fine-grained music scores are obtained through a  
complicated data collection pipeline, which can be mainly divided into three major steps\citep{wang2022opencpop,zhangm4singer}: 
1) phoneme annotation step, where the duration of each phoneme is first extracted from singing through Montreal Forced Aligner\footnote{https://github.com/MontrealCorpusTools/Montreal-Forced-
Aligner} and then further manually annotated to acquire more accurate phoneme boundaries.
2) note annotation step, where preliminary notes are either created by Logic Pro\citep{wang2022opencpop} or collected through the word-level average of extracted F0\citep{zhangm4singer} and then note durations are manually adjusted to the boundaries of vowel phonemes. 3) silence annotation step, where the silence part is annotated as silence (SP) or aspirate (AP). 

These steps, especially the first and second steps, require arduous and professional manual annotation\cite{zhangm4singer}, which hinders large-scale SVS data collection. Furthermore, since the manual adjustment in the second step destroys the regularity of note durations, fine-grained music scores have a noteworthy difference from realistic music scores (see Figure \ref{fig_data}), which inhibits human composers from employing SVS methods for composing.

The primary rationale for adopting such a time-consuming and laborious data collection pipeline is twofold:
1) Existing methods require phoneme-level hard-alignment for duration training. Due to the difficulty in determining phoneme boundaries\citep{ren2021portaspeech}, complex manual annotation (phoneme annotation step) is necessary in order to prevent the negative effects of incorrect alignment on model training.
2) Existing methods require a pre-defined phoneme-level mel-note alignment for training and inference. 
Since one vowel phoneme may correspond to multiple notes\citep{wang2022opencpop},  existing methods directly repeat this phoneme to conform with notes, 
which requires note boundaries to be aligned with the boundary of each vowel phoneme. However, even professional singers can
hardly sing fully conformed to the music score\citep{zhangm4singer}, so the note annotation step has to be performed.

To tackle these challenges, we introduce Realistic-Music-Score Singer (RMSSinger), the first RMS-SVS method, which utilizes realistic music scores with different note types (grace, slur, rest, etc.,) for training and inference, alleviating most manual annotations. To alleviate the tedious annotation in the phoneme annotation step, we propose word-level positional attention with word-level hard-alignment and positional attention to avoid the difficulty of determining exact phoneme boundaries. To avoid the note annotation step, we propose the word-level learned Gaussian upsampler to learn the word-level mel-note alignment in training and avoid the phoneme-level mel-note alignment.
Furthermore, existing methods mainly adopt simple L1 or L2 loss for pitch modeling, which results in the degradation of expressiveness. To achieve expressive pitch prediction, we propose the first diffusion-based pitch generation method. Due to the existence of both continuous parts (F0) and categorical parts (UV) in pitch contours, we propose the pitch diffusion model (P-DDPM), which models categorical UV and continuous F0 in a single model. Extensive experiments on our collected datasets demonstrate the efficiency of our proposed word-level framework (word-level positional attention and word-level learned Gaussian upsampler) and P-DDPM. The main contributions of this work are summarized as follows:
\begin{itemize}
    \item We propose the first realistic-music-score-based singing voice synthesis method RMSSinger, which alleviates tedious manual annotation in the current SVS data collection pipeline and achieve high-quality singing voice synthesis given realistic music scores.
    \item We propose the word-level positional attention and the word-level learned Gaussian upsampler to model lyrics and notes on the word level and avoid phoneme duration annotation and phoneme-level mel-note alignment.
    \item We propose the first diffusion-based pitch generation model (P-DDPM), which models the continuous F0 and categorical UV in a single model and improves the expressiveness of pitch modeling.
    \item Extensive experiments demonstrate the performance of our proposed method.
\end{itemize}

\section{Related Works}
Singing Voice Synthesis (SVS) aims to generate high-quality singing conditioned on given music scores. With the development of deep learning, SVS has achieved great progress in the network structure and the singing corpus construction. XiaoiceSing\citep{lu2020xiaoicesing} adopts the non-autoregressive acoustic model inspired by FastSpeech\citep{ren2019fastspeech}. ByteSing\citep{gu2021bytesing} is designed based on the auto-regressive Tacotron-like\citep{wang2017tacotron} architecture. DeepSinger\citep{ren2020deepsinger} builds a singing corpus by mining singing data from websites and proposes the singing model based on the feed-forward transformer\citep{ren2019fastspeech}.
More recently, Opencpop\citep{wang2022opencpop} publish a single-singer Chinese song corpus with manually-annotated fine-grained music scores and propose a Conformer-based\citep{gulati2020conformer} SVS method.
WeSinger\citep{zhang2022wesinger} adopts a Transformer-like acoustic model and an LPCNet neural vocoder. ViSinger\citep{zhang2022visinger} employs the VITS\citep{kim2021conditional} architecture for end-to-end SVS and introduces an F0 predictor to guide the prior network. DiffSinger\citep{liu2022diffsinger} introduces the diffusion-based\citep{ho2020denoising} decoder for the high-quality mel-spectrogram generation and proposed the shallow diffusion mechanism for faster inference.
M4Singer\citep{zhangm4singer} further publishes a multi-style, multi-singer Chinese song corpus with manually-annotated fine-grained music scores.

\section{Diffusion Models}
Diffusion models\citep{sohl2015deep,ho2020denoising,song2020denoising} are a paradigm of generative methods that aim to approximate the end-point distribution (target distribution) of a Markow chain and have achieved impressive results in benchmark generative tasks\citep{dhariwal2021diffusion}. Diffusion models consist of two processes:

\noindent \textbf{Diffusion Process}
The diffusion process gradually perturbs data $x_0 \sim q(x_0)$ to pure noise with a Markov chain according to the variance schedule $\beta_1,...,\beta_T$:
\begin{equation}
  \begin{aligned}
    q(x_{1:T}|x_0) = \prod_{t=1}^Tq(x_t|x_{t-1}).
  \end{aligned}
\end{equation}
\noindent \textbf{Reverse Process}
The reverse process gradually denoises the latent variable $x_T \sim p(x_T)$ to the corresponding real data sample $x_0$:
\begin{equation}
  \begin{aligned}
    p_{\theta}(x_{0:T}) = p(x_T)\prod_{t=1}^Tp_\theta(x_{t-1}|x_{t}),
  \end{aligned}
\end{equation}
where $p_{\theta}(x_{t-1}|x_{t})$ are parameterized with a neural network and learned by optimizing the usual variational bound on negative log-likelihood:
\begin{equation}
\label{nll}
  \begin{aligned}
    &E[-logp_\theta(x_0)] \leq \\ 
    &E_q[-logp(x_T)-
    \sum_{t \geq 1}log\frac{p_\theta(x_{t-1}|x_{t})}{q(x_t|x_{t-1})}] = \mathcal{L},\\
    &\mathcal{L}_{t-1} = \mathcal{D}_{KL}(q(x_{t-1}|x_t,x_0)|| p_\theta(x_{t-1}|x_{t}))
  \end{aligned}
\end{equation}

With different perturbation transition $q(x_t|x_{t-1})$ used, different diffusion models are defined:

\noindent \textbf{Gaussian Diffusion:} Gaussian diffusion\citep{ho2020denoising,nichol2021improved} is utilized in continuous data domains. Gaussian diffusion adopts the Gaussian noise for perturbation:
\begin{equation}
\label{eq_gauss_inp}
  \begin{aligned}
   q(x_t|x_{t-1}) &= \mathcal{N}(x_t;\sqrt{1-\beta_t}x_{t-1},\beta_t I),\\
   p(x_{t-1}|x_t) &= \mathcal{N}(x_{t-1};\mu_{\theta}(x_t, t),\Sigma_{\theta}(x_t, t)).
  \end{aligned}
\end{equation}
With the parameterization introduced in \citep{ho2020denoising}, Equation \ref{nll} can be further simplified and finally optimized with:
\begin{equation}
\label{eq_gauss_out}
  \begin{aligned}
   E_{x_0,\epsilon}[\frac{\beta_t^2}{2\sigma_t^2\alpha_t(1-\bar \alpha_t)}||\epsilon-\epsilon_\theta(x_t,t)||],
  \end{aligned}
\end{equation}
where $\alpha_t = 1-\beta_t,\bar \alpha_t=\prod_{s=1}^t\alpha_s$. The neural network is trained to predict the "noise" $\epsilon$ from noisy input $x_t$ given timestep $t$.
Gaussian diffusion has been widely utilized for image generation\citep{nichol2021improved,dhariwal2021diffusion} and audio generation\citep{jeong2021diff,huang2022fastdiff}.

\noindent \textbf{Multinomial Diffusion:} Multinomial diffusion \citep{hoogeboom2021argmax} is utilized in discrete data domains, which replaces Gaussian noise with random walking on discrete data space. The diffusion process can then be defined as:
\begin{equation}
\label{eq_md_f}
  \begin{aligned}
   &q(x_t|x_{t-1}) = \mathcal{C}(x_t|(1-\beta_t)x_{t-1}+\beta_t/K),\\
   &q(x_t|x_0) = \mathcal{C}(x_t|\bar \alpha_t x_0 + (1-\bar \alpha_t)/K)
  \end{aligned}
\end{equation}
where $\mathcal{C}$ denotes a categorical distribution with probability parameters, $x_t \sim \{0,1\}^K$, $\beta_t$ denotes the chance of
resampling a category uniformly, 
and $\alpha_t = 1-\beta_t,\bar \alpha_t=\prod_{s=1}^t\alpha_s$.
Using Equation \ref{eq_md_f}, we can compute the categorical posterior:
\begin{equation}
\label{eq_md_r}
  \begin{aligned}
   &q(x_{t-1}|x_t, x_0) = \mathcal{C}(x_{t-1}|\theta_{post}(x_t, x_0)),\\ &\theta_{post}(x_t, x_0)=\tilde{\theta}/\sum_{k=1}^K\tilde{\theta_k},\\
   &\tilde{\theta} = [\alpha_tx_t + (1-\alpha_t)/K] \odot [\bar \alpha_{t-1}x_0+(1-\bar \alpha_{t-1})/K],\\
  \end{aligned}
\end{equation}
With the parameterization proposed in \citep{hoogeboom2021argmax}, $p(x_{t-1}|x_t) = \mathcal{C}(x_{t-1}|\theta_{post}(x_t, \hat{x_0}))$ is utilized to approximate $q(x_{t-1}|x_t, x_0)$. And the neural network is trained to approximate $\hat{x_0}$ from noisy sample $x_t$ given timestep $t$.

Though widely utilized in many data domains, diffusion models have never been utilized for pitch modeling. Furthermore, due to the existence of continuous F0 parts and discrete UV parts in pitch contours\citep{wang2018autoregressive}, neither Gaussian diffusion nor multinomial diffusion alone can deal with pitch modeling. In this paper, we propose the first diffusion-based pitch modeling (P-DDPM), which incorporates Gaussian diffusion and multinomial diffusion in a single model and achieves better pitch modeling.

\section{Methdology}
\begin{figure*}
\begin{minipage}[b]{.4\linewidth}
    \centering
    \subfloat[][overall architecture]{\label{fig_overall}\includegraphics[width=0.9\linewidth]{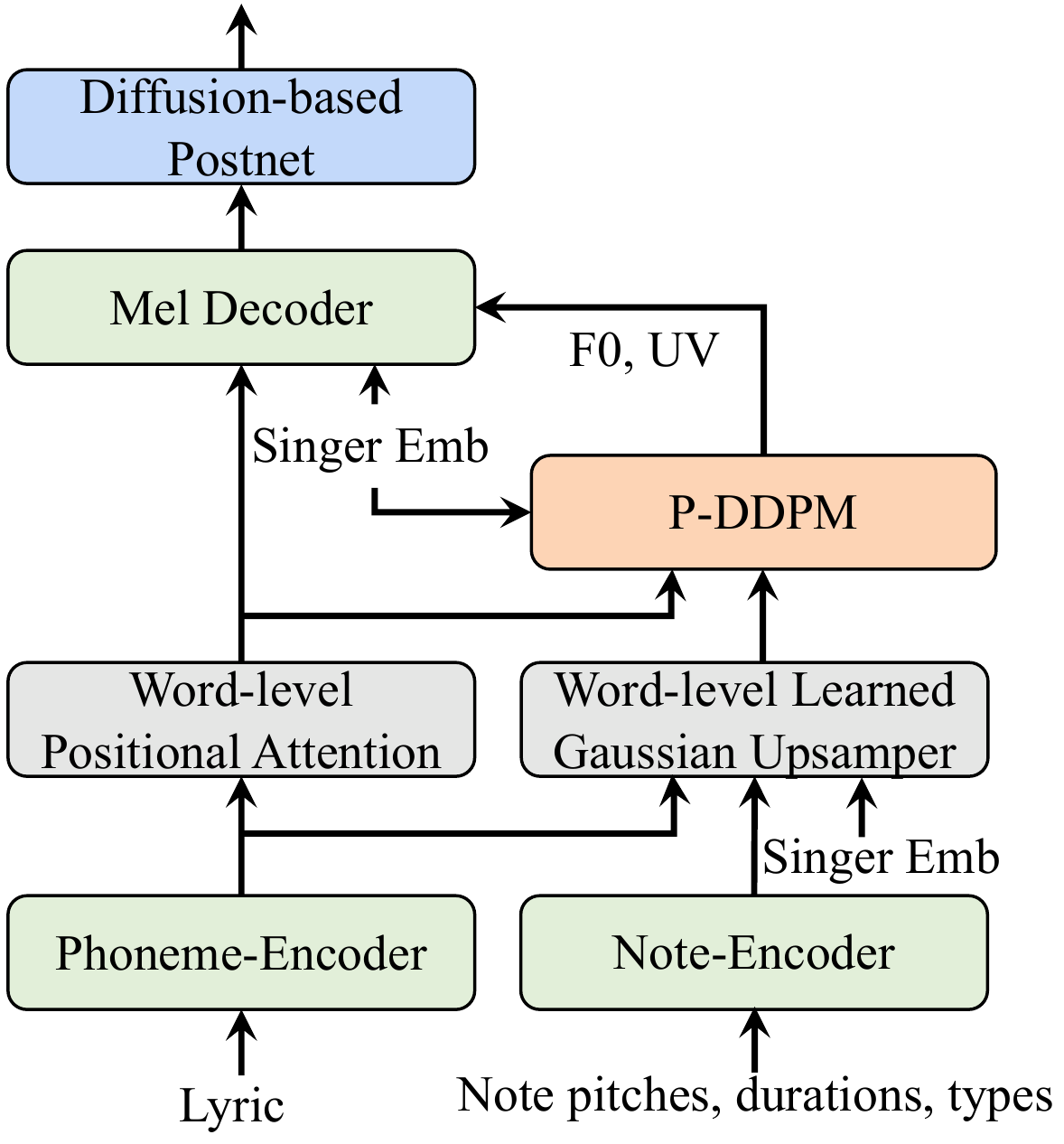}}
\end{minipage} 
\medskip
\begin{minipage}[b]{.25\linewidth}
    \centering
    \subfloat[][word-level learned Gaussian upsampler]{\label{fig_gaussian}\includegraphics[width=1\linewidth]{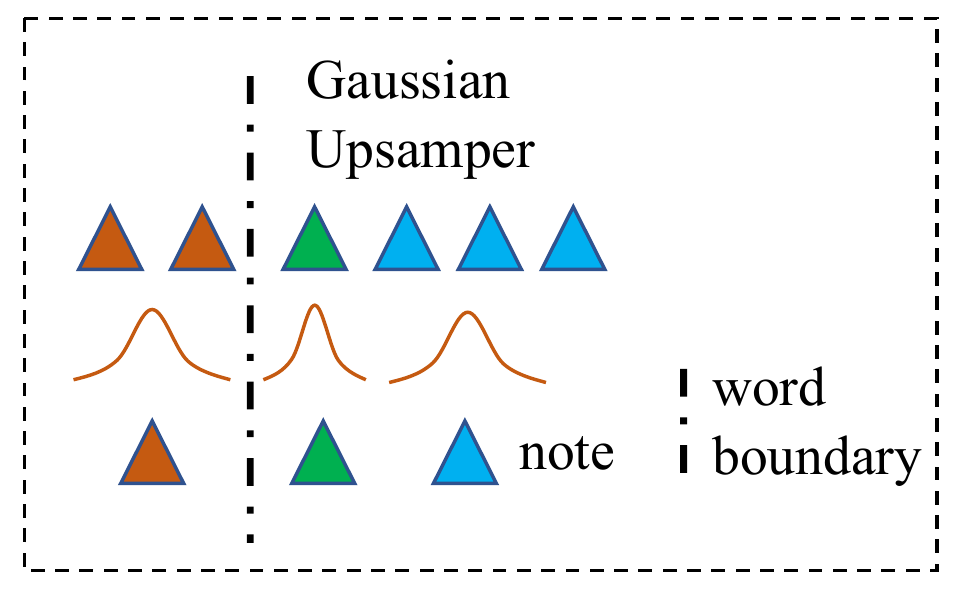}}

    \subfloat[][word-level positional attention]{\label{fig_pa}\includegraphics[width=1\linewidth]{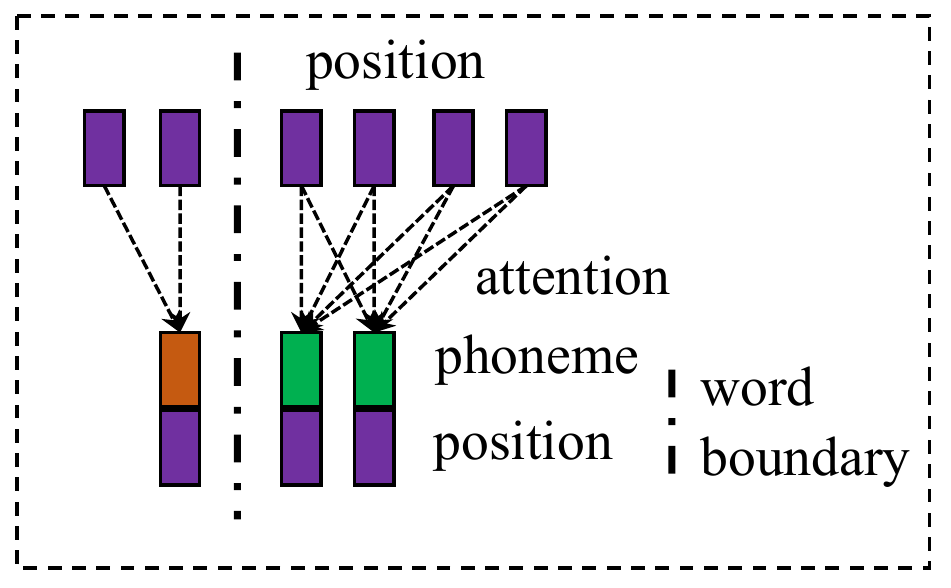}}
\end{minipage}
\medskip
\begin{minipage}[b]{.35\linewidth}
    \centering
    \subfloat[][P-DDPM]{\label{fig_pddpm}\includegraphics[width=0.9\linewidth]{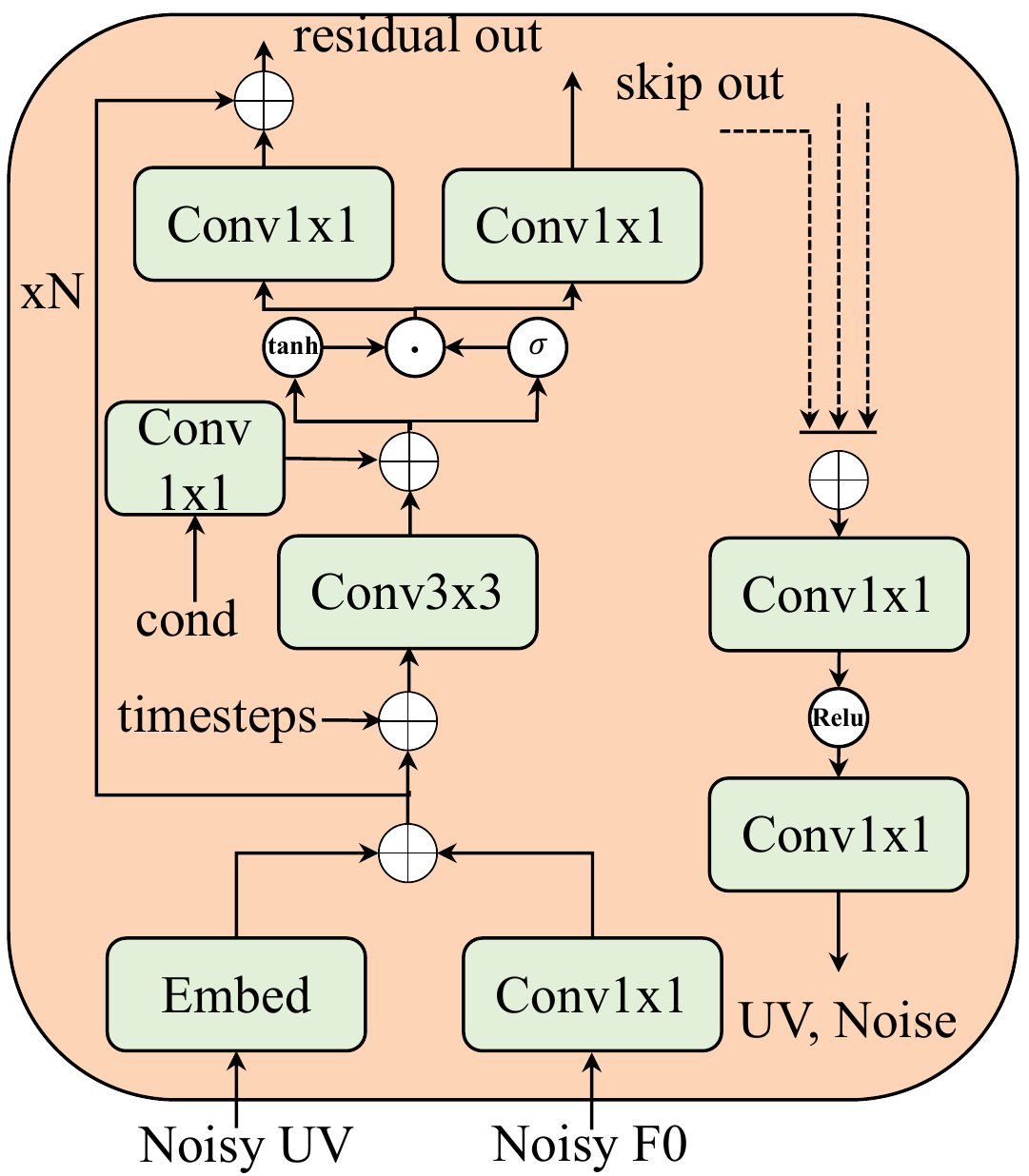}}
\end{minipage}
\vspace{-0.4cm}
\caption{The overall architecture of our proposed RMSSinger is shown in subfigure (a). Lyrics are first encoded through the phoneme encoder and then expanded through the word-level positional attention in subfigure (c). Notes are encoded through the note encoder and then expanded through the word-level learned Gaussian upsampler in subfigure (b). The expanded note feature, expanded lyric feature and singer embedding are summed and used as the condition of P-DDPM in subfigure (d).}
\label{fig:Genelecs}
\vspace{-0.4cm}
\end{figure*}

\subsection{Overview}
In this section, we introduce the overall architecture of our proposed RMSSinger. As shown in Figure \ref{fig_overall}, RMSSinger is built on one of the most popular non-autoregressive TTS models FastSpeech2\citep{ren2020fastspeech}. Lyrics are encoded through the phoneme encoder and then aligned to the lengths of mel-spectrogram through the word-level positional attention layer (Section \ref{sec_pa})to obtain the expanded lyric feature. Next, we utilize the note encoder to encode note pitches, note durations, and note types (rest, slur, grace, etc.) and adopt the word-level learned Gaussian upsampler (Section \ref{sec_gu}) for word-level mel-note alignment learning to obtain the expanded note feature. The timbre information of different singers is embedded to obtain the singer embedding.
Then, the expanded lyric feature, expanded note feature and singer embedding are summed as the pitch decoder input. The pitch diffusion model (P-DDPM) (Section \ref{sec_pddpm}) utilizes the pitch decoder input as the condition to generate pitches (F0 and UV). Similar to \citep{ren2020fastspeech}, we obtain the pitch embedding through F0 and UV. 
Next, the expanded lyric feature, pitch embedding and singer embedding are summed as the input of the mel decoder. Finally, to further improve the quality of the synthesized mel-spectrogram, we introduce a diffusion-based post-net (Section \ref{sec_post}) to refine the coarse outputs of the mel decoder.

\subsection{Encoder}
In this subsection, we introduce the phoneme encoder and the note encoder utilized in RMSSinger. The phoneme encoder takes the phoneme sequence as input and outputs the phoneme feature $\mathcal{H}$. We also perform the word-pooling on $\mathcal{H}$ to obtain the word-level feature $\mathcal{H}_w$.
The architecture of the phoneme encoder is comprised of a series of Feed-Forward Transformer Blocks \citep{vaswani2017attention}, which have proven the effectiveness of long sequences modeling and linguistic information extraction in TTS methods. The input of the note encoder is the realistic music score. As there exist different types of information in music scores, the note encoder includes an embedding layer for note pitches, an embedding layer for note types (rest, slur, grace, etc.), and a linear projection layer for note durations. All information types are summed as the note feature $\mathcal{H}_n$.

\subsection{Word-level Learned Gaussian Upsampler}
\label{sec_gu}
One of the key challenges of SVS is the alignment between word-level mel-spectrogram and notes, that is the actual length\footnote{The number of the mel-frames} of each note.
Though the note duration on music scores provides a preliminary estimate of the actual length, even professional singers cannot precisely conform to the music score. Therefore, previous SVS methods manually adjust the note duration to the phoneme boundary, which not only requires time-consuming annotation from experts but also destroys the regularity of the note duration. 

The key idea of the proposed word-level learned Gaussian upsampler (see figure \ref{fig_gaussian}), inspired by \citep{donahue2020end}, is to learn the word-level mel-note alignment in training. 
Given the note feature $\mathcal{H}_n$, the word-level feature $\mathcal{H}_w$ and the singer embedding $s$, we expand $\mathcal{H}_w$ to the note-level $\mathcal{H}_{wn}$. Next we
predict the actual length of each note:
\begin{equation}
  \begin{aligned}
   &\mathcal{H}_a = \mathcal{H}_n + \mathcal{H}_{wn} + s, \\
   &l_n = f(\mathcal{H}_a),
  \end{aligned}
\end{equation}
with a neural network $f$. The neural network consists of a stack of 1D-convolution, Relu, and layer normalization. We use a linear projection with ReLU nonlinearity at the output to make $l_n$ non-negative, which ensures the monotonicity and none of the notes can be ignored.
Then we upsample the note feature to its corresponding actual length. We introduce the Gaussian distribution to make the upsampling process differentiable and learnable. To be specific, given the predicted actual length, we can find the end position of each note $e_n = \sum_{m=1}^{n}l_m$, and then the center position of each note $c_n = e_n - \frac{1}{2}l_n$. We place a Gaussian distribution with fixed deviation $\sigma$ at the center $c_n$ of the output segment corresponding to the note $n$. Then we can define:
\begin{equation}
  \begin{aligned}
    w_t^n = \frac{exp(-\frac{(t-c_n)^2}{2\sigma^2})}{\sum_m exp(-\frac{(t-c_m)^2}{2\sigma^2})}, t \in (0, T)
  \end{aligned}
\end{equation}
where $T$ denotes the length of the mel-spectrogram and $w_t^n$ represents the weight of each note for the output position $t$. And finally, the expanded note feature at position $t$ can be calculated as $a_t = \sum_{n}w_t^n\mathcal{H}_n$. We highlight that 1) when calculating $a_t$, we only consider the contribution from the same word, that is if position $t$ belongs to the range of word $i$ and note $n$ corresponds to the word $j$ and $i \neq j$, then $w_t^n = 0$. 
2) During training, we use the ground-truth duration of each word to determine the range of position $t$ and avoid expensive DTW calculation. 2) During inference, we use the sum of the predicted actual length of notes $\sum_m l_m, m \in word_i$ as the predicted duration of $word_i$. 3) We use the ground-truth word duration to constrain the prediction of the actual length, which is computed as:
\begin{equation}
  \begin{aligned}
    \mathcal{L}_d = || \sum_m l_m - dur_i ||,
  \end{aligned}
\end{equation}
where $dur_i$ denotes the ground-truth duration of $word_i$.

\subsection{Word-level Positional Attention}
\label{sec_pa}
To align the lyric features (outputs of the phoneme encoder) to the lengths of the mel-spectrogram, previous SVS methods mainly adopt the duration predictor to predict the number of frames of each phoneme. Due to the complex articulation of each phoneme in singing, these methods have to use manually-annotated phoneme duration for training, which increases the cost of data collection. Note that most music scores are word-level and word boundaries are much easier to be determined, inspired by \citep{ren2021portaspeech,miao2020flow}, we propose the word-level positional attention (see figure \ref{fig_pa}), which avoids the annotation of phoneme duration.
To be specific, given the output of the phoneme encoder $\mathcal{H}$, let 
the word-level phoneme positional encoding  which represents the position of each phoneme in a word be $\mathcal{P}_{ph}$, and let the word-level mel-spectrogram positional encoding which denotes the position of each frame in a word be $\mathcal{P}_m$, we introduce the position-to-phoneme attention:
\begin{equation}
  \begin{aligned}
    &\mathcal{H}_k = W(cat(\mathcal{H},\mathcal{P}_{ph})),\\
    &\mathcal{H}_{epd} = Softmax(\frac{\mathcal{P}_m\mathcal{H}_k^T}{\sqrt{d}})\mathcal{H}^T,
  \end{aligned}
\end{equation}
where $W$ represents a linear projection, and $\mathcal{H}_{epd}$ represents the expanded lyric-feature. During training, we use the ground-truth word durations to obtain word-level mel-spectrogram positional encoding. During inference, we use the predicted word duration $\sum_m l_m$ introduced in subsection \ref{sec_gu}.

\subsection{Pitch Diffusion Model}
\label{sec_pddpm}
To generate the pitch contours, previous methods mainly adopt a pitch predictor which predicts the continuous fundamental frequency (F0) and the discrete unvoiced label (UV). The pitch predictor is constrained with simple L1 or L2 loss for F0 and cross-entropy loss for UV. However, due to the complicated pitch variation of the singing voice, the simple pitch predictor fails to model the variance, resulting in degraded expressiveness. To tackle this challenge, we propose the first pitch diffusion model (P-DDPM) (see Figure \ref{fig_pddpm}), which incorporates both the Gaussian diffusion and multinomial diffusion to generate F0 and UV. 
During the diffusion process, the Gaussian noise (see Equation \ref{eq_gauss_inp}) and random resampling (see Equation \ref{eq_md_f}) are used to perturb the continuous F0 (represented by $x$) and the discrete UV labels (represented by $y$) correspondingly: 
\begin{equation}
  \begin{aligned}
    &q(x_t|x_{t-1}) = \mathcal{N}(x_t;\sqrt{1-\beta_t}x_{t-1},\beta_t I),\\
    &q(y_t|y_{t-1}) = \mathcal{C}(y_{t}|(1-\beta_t)y_{t-1}+\beta_t/K).
  \end{aligned}
\end{equation}
During the reverse process, 
following Equation \ref{eq_gauss_out} and Equation \ref{eq_md_r}, the neural network is utilized to predict the corresponding $\epsilon_{\theta}(x_t,t)$ and $\hat{y_0}$.
We adopt a non-causal WaveNet \citep{oord2016wavenet} architecture as our denoiser, which has proven to be effective in modeling sequential data. We design a 1x1 convolution layer for the continuous F0 and an embedding layer for the discrete UV label in order to perform Gaussian F0 diffusion and multinomial UV diffusion in a single model. The neural network is optimized through the corresponding Gaussian diffusion loss $\mathcal{L}_{gdiff}$ and multinomial diffusion loss $\mathcal{L}_{mdiff}$.

\subsection{Decoder}
In this subsection, we introduce the mel decoder utilized in RMSSinger. The mel decoder takes the expanded lyric feature, singer embedding and pitch embedding as input and outputs the coarse mel-spectrogram. Following previous speech synthesis methods \citep{huang2022generspeech,he2022flow}, we use a stack of Feed-Forward Transformer blocks as the architecture and use the L1 loss function to optimize the mel decoder:
\begin{equation}
  \begin{aligned}
   \mathcal{L}_{mel} = ||mel_p - mel_g||,
  \end{aligned}
\end{equation}
where $mel_p$ denotes the predicted coarse mel-spectrogram and $mel_g$ denotes the ground-truth mel-spectrogram. 

\subsection{Diffusion-Based Post-Net}
\label{sec_post}
To achieve high-quality singing voice synthesis, we have to capture the rich and highly dynamic variation in the singing voice. However, the widely-applied transformer-based decoder (mel decoder) is difficult to generate detailed mel-spectrogram samples\citep{ren2022revisiting}. To further improve the quality of generated samples, we introduce the diffusion-based post-net, which converts the coarse outputs of the mel decoder into fine-grained ones. In detail, we use the coarse outputs as the condition of the diffusion model for training and inference. We use the Gaussian diffusion loss $\mathcal{L}_{post}$ similar to the previous diffusion-based TTS method \citep{jeong2021diff} to optimize the diffusion-based postnet.

\subsection{Training Pipeline}
There are two training stages for RMSSinger: during the first stage, we optimize the whole model except the diffusion-based postnet by minimizing the following loss function:
\begin{equation}
  \begin{aligned}
    \mathcal{L}_1 = \mathcal{L}_{gdiff} + \mathcal{L}_{mdiff} + \mathcal{L}_d + \mathcal{L}_{mel}
  \end{aligned}
\end{equation}
We obtain coarse mel-spectrogram after the first stage of training. In the second training stage, we freeze the whole model except the diffusion-based postnet and only optimize the diffusion-based postnet by minimizing  $\mathcal{L}_{post}$.

\begin{table*}[h]
\centering
\begin{tabular}{l|l|l|l|l|l}
\hline
Method & F0RMSE $\downarrow$ & VDE $\downarrow$ & MCD $\downarrow$ & MOS-P $\uparrow$ & MOS-Q $\uparrow$\\
\hline
GT & - & - & - & 4.55 $\pm$ 0.04 & 4.58 $\pm$ 0.03
\\
GT(vocoder)  & 3.77 & 0.020 & 1.33 & 4.10 $\pm$ 0.04 & 4.09 $\pm$ 0.05 \\
\hline
FFTSinger\citep{zhangm4singer} & 14.0 & 0.092 & 3.52 & 3.57 $\pm$ 0.08 & 3.46 $\pm$ 0.07 
\\
DiffSinger\citep{liu2022diffsinger} & 12.4 & 0.077 & 3.43 & 3.63 $\pm$ 0.07 & 3.79 $\pm$ 0.07
\\
\hline
RMSSinger (ours) & \textbf{12.2} & \textbf{0.069} & \textbf{3.42} & \textbf{3.77} $\pm$ \textbf{0.05} & \textbf{3.84} $\pm$ \textbf{0.06}
\\
\hline
\end{tabular}
\caption{\label{baseline}
Performance comparison with different methods. We use F0RMSE, VDE and MCD for objective evaluation. And we use MOS-P and MOS-Q for subjective measurement.
}
\end{table*}

\begin{figure*}
\begin{minipage}[b]{.24\linewidth}
    \centering
    \subfloat[][ground-truth]{\includegraphics[width=1.0\linewidth]{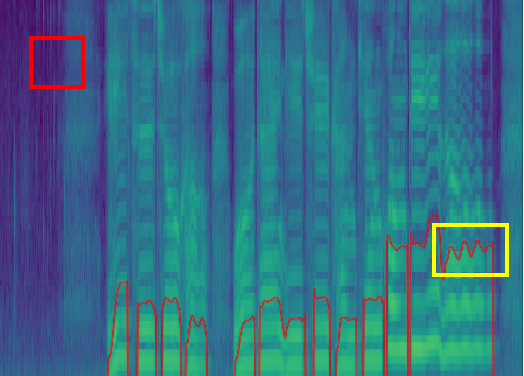}}
\end{minipage} 
\begin{minipage}[b]{.24\linewidth}
    \centering
    \subfloat[][FFTSinger]{\includegraphics[width=1.0\linewidth]{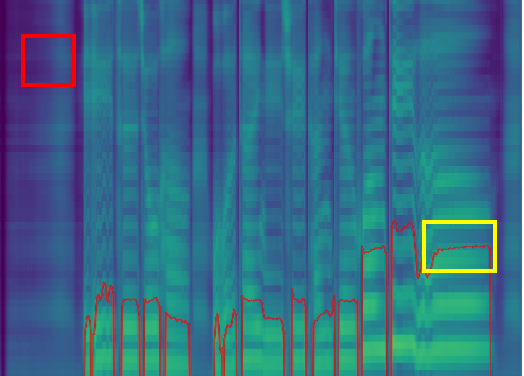}}
\end{minipage}
\begin{minipage}[b]{.24\linewidth}
    \centering
    \subfloat[][DiffSinger]{\includegraphics[width=1.0\linewidth]{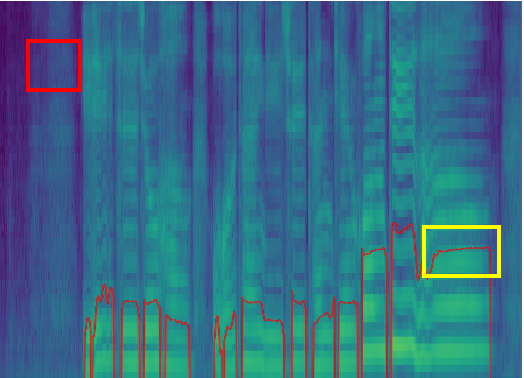}}
\end{minipage} 
\begin{minipage}[b]{.24\linewidth}
    \centering
    \subfloat[][RMSSinger]{\includegraphics[width=1.0\linewidth]{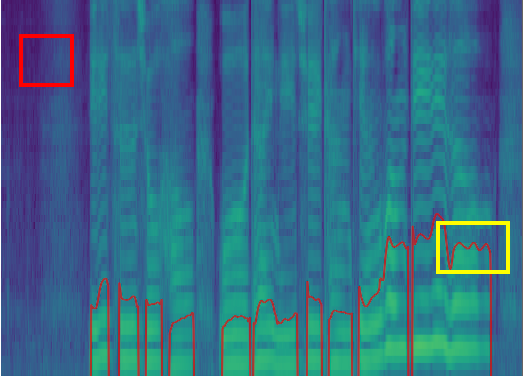}}
\end{minipage}
\caption{Visualization of the pitch contour and mel-spectrograms of ground-truth and different methods.}
\label{fig:main}
\end{figure*}

\section{Experiments}
\subsection{Experimental Setup}
In this section, we first describe our collected dataset for RMS-SVS, and then introduce the implementation details of our proposed RMSSinger. Finally, we explain the training and evaluation details utilized in this paper.

\noindent \textbf{Dataset} Currently, there are no public SVS datasets providing realistic music scores, so we collect and annotate a high-quality Chinese song corpus (about 12 hours in total) with realistic music scores. Professional singers are recruited to sing conforming to these realistic music scores. They are paid based on their singing time.
Next, word durations are extracted through an external speech-text aligner and then manually finetuned. Since we do not need fine-grained phoneme durations, the finetune process requires much less effort. Finally, we annotate the silence and aspirate parts since these parts are not provided in most realistic music scores. All audio files are recorded in a professional recording studio, which guarantees the high quality of our dataset. All audios are sampled as 48000 Hz with 24-bit quantization, and we randomly select one song from each singer for the testing.

\noindent \textbf{Implementation Details}
We convert Chinese lyrics into phonemes through pypinyin. 
We extract mel-spectrogram from raw waveforms and set the sample rate to 24000Hz, the window size to 512, the hop size to 128, and the number of mel bins to 80. In the phoneme encoder and the mel decoder, we adopt a similar setting as that in FastSpeech2\citep{ren2020fastspeech}. In the word-level learned Gaussian upsampler, the kernel size of 1D-convolution is set to 5 and the hidden channel is set to 256. In the word-level positional attention, we set the number of attention heads to 1. In the P-DDPM, we set the number of convolution layers to 12, the kernel size to 3, the residual channel to 192 and hidden channel to 256. And we also set the total number of diffusion steps to 100 and adopt the linear $\beta$ schedule from 0.0001 to 0.06. The diffusion post-net has a similar architecture and $\beta$ schedule except that the number of convolution layers is set to 20, and the residual channel is set to 256. More details can be found in Appendix \ref{sec:app_model}.

\noindent \textbf{Evaluation Details} 
In our experiments, we employ objective and subjective evaluation metrics to evaluate the pitch modeling and the audio quality of generated samples.
For the objective evaluation, we utilize F0 Root Mean Square Error(F0RMSE) to measure the accuracy of F0 prediction and Voice Decision Error(VDE) to measure the accuracy of UV prediction. We use Mean Cepstral Distortion (MCD) for audio quality measurement.
For the subjective evaluation, we use Mean Opinion Score (MOS) for main results and Comparision Mean Opinion Score (CMOS) for ablations. For a more detailed examination, we score MOS-P/CMOS-P and MOS-Q/CMOS-Q corresponding to the MOS/CMOS of pitch modeling and audio quality. We utilize the HiFi-GAN\citep{kong2020hifi} vocoder published in DiffSinger\citep{liu2022diffsinger} for all experiments. More details can be found in Appendix \ref{sec:app_exp}.

\begin{figure*}
\begin{minipage}[b]{.24\linewidth}
    \centering
    \subfloat[][ground-truth]{\includegraphics[width=1.0\linewidth]{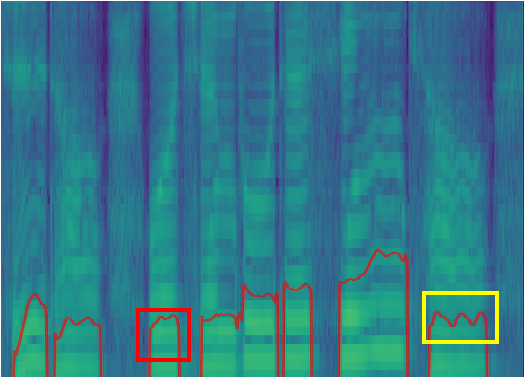}}
\end{minipage} 
\begin{minipage}[b]{.24\linewidth}
    \centering
    \subfloat[][RMSSinger]{\includegraphics[width=1.0\linewidth]{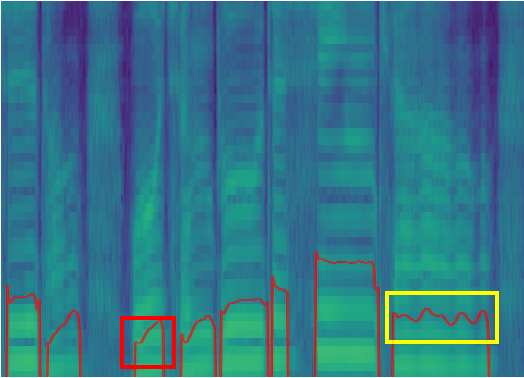}}
\end{minipage}
\begin{minipage}[b]{.24\linewidth}
    \centering
    \subfloat[][w/o UV diffusion]{\includegraphics[width=1.0\linewidth]{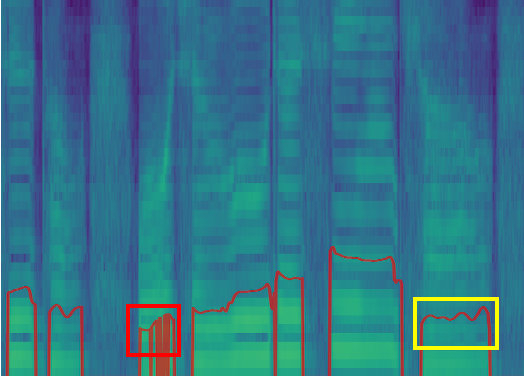}}
\end{minipage} 
\begin{minipage}[b]{.24\linewidth}
    \centering
    \subfloat[][w/o F0 diffusion]{\includegraphics[width=1.0\linewidth]{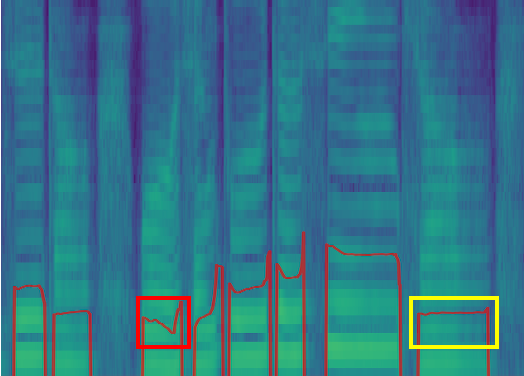}}
\end{minipage}
\caption{Visualization of ground truth and generated results of RMSSinger and different ablations}
\label{fig:ablation}
\end{figure*}

\subsection{Main Results}
In this subsection, we conduct extensive experiments to compare the performance of RMSSinger with other baselines.
Since RMS-SVS is a new task, none of the existing methods can handle it. Therefore, we implement several representative and state-of-art SVS methods\citep{zhangm4singer,liu2022diffsinger} with our proposed word-level framework to handle realistic music scores. 
Specifically, FFTSinger\citep{zhangm4singer} adopts a similar architecture to FastSpeech2\citep{ren2020fastspeech}, which uses MSE loss for F0 training and binary cross entropy loss for UV training. Besides, FFTSinger adopts the FFT decoder and uses the L1 loss for mel-spectrogram reconstruction.
DiffSinger\citep{liu2022diffsinger} uses the same pitch modeling method but replaces the FFT decoder in FFTSinger with the diffusion-based decoder and uses the Gaussian diffusion for mel-spectrogram training.

The main results are shown in Table \ref{baseline}. From the objective and subjective results, we can see that 1) most methods achieve promising results, which illustrates the feasibility of RMS-SVS and the effectiveness of our proposed word-level framework. 2) RMSSinger achieves better results on F0RMSE, VDE, and MOS-P, which demonstrates our proposed P-DDPM can improve both the F0 and UV modeling and improve the expressiveness. 3) RMSSinger and DiffSinger achieve similar results on MCD and MOS-Q, which is because the diffusion-based postnet of RMSSinger and the diffusion decoder of DiffSinger have a similar architecture.
4) RMSSinger and DiffSinger outperform FFTSinger in terms of audio quality by a large margin due to the existence of mel-spectrogram over-smooth in FFTSinger.

We then visualize the mel-spectrogram and pitch contour generated by different methods in Figure \ref{fig:main} to show the difference among different methods more intuitively.
We can find that 1) RMSSinger can generate more natural pitch contours especially in the vibrato part (yellow box region), which demonstrates our method can achieve better pitch modeling. 2) RMSSinger and DiffSinger can generate more detailed mel-spectrogram, and alleviate the mel-spectrogram over-smooth (see red box region), which explains the higher audio quality.

\begin{table}[h]
\centering
\scalebox{0.94}{
\begin{tabular}{l|l|l|l}
\hline
Method & F0RMSE & VDE & CMOS-P \\
\hline
Full model & \textbf{12.2} & \textbf{0.069}  & \textbf{0.0} 
\\
w/o UV diffusion & 12.6 & 0.080  & -0.58
\\
w/o F0 diffusion & 12.8 & 0.070   & -0.50
\\
\hline
\end{tabular}}
\caption{\label{ablation_p}
Ablation studies on the effect of P-DDPM.
}
\end{table}

\begin{table}[h]
\centering
\begin{tabular}{l|l|l}
\hline
Method & MCD &CMOS-Q\\
\hline
Full model & \textbf{3.42}  & \textbf{0.0} 
\\
w/o diffusion postnet & 3.48  & -0.91 
\\
\hline
\end{tabular}
\caption{\label{ablation_q}
Ablation studies on the postnet.
}
\end{table}

\subsection{Ablation Studies}
In this subsection, we conduct a series of ablation studies to investigate the effect of key components in our RMSSinger. 

\noindent \textbf{P-DDPM} To evaluate the effectiveness of our proposed P-DDPM, we design two ablations. In the first ablation,
we remove the UV diffusion from P-DDPM and we use a Transformer-based UV predictor, which is constrained by binary cross-entropy loss. In the second ablation,
we remove the F0 diffusion from P-DDPM and use a Transformer-based F0 predictor, which is constrained by L1 loss. The results can be found in Table \ref{ablation_p}. We can see that 1) in the first ablation, VDE increases significantly and CMOS-P degrades, which demonstrates that the UV diffusion in P-DDPM contributes to better UV modeling;  
2) in the second ablation, F0RMSE increases, VDE nearly holds the same and CMOS-P degrades which demonstrates the F0 diffusion is essential to natural F0 modeling. 

We also visualize the pitch contours of different ablations in Figure \ref{fig:ablation}. We can find that 1) without UV diffusion, there exists unpleasant UV errors (see the red box). 2) without F0 diffusion, the model can not generate natural F0, especially in the vibrato part (yellow box region).
\begin{figure}
\begin{minipage}[b]{.48\linewidth}
    \centering
    \subfloat[][RMSSinger]{\includegraphics[width=1.0\linewidth]{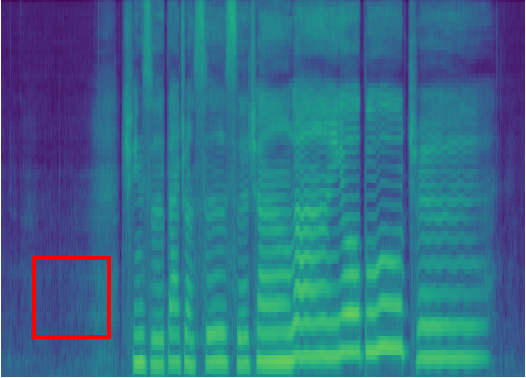}}
\end{minipage}
\begin{minipage}[b]{.48\linewidth}
    \centering
    \subfloat[][w/o postnet]{\includegraphics[width=1.0\linewidth]{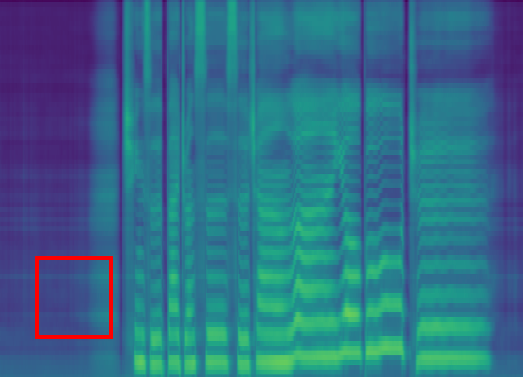}}
\end{minipage} 
\caption{Visualization of generated results of RMSSinger and the ablation}
\label{fig:ablation1}
\end{figure}

\noindent \textbf{Postnet} To evaluate the effectiveness of our proposed diffusion-based postnet, we remove the postnet and utilize the output of the mel decoder as the generated samples. The results can be found in Table \ref{ablation_q}. We can find that MCD increases and CMOS-Q degrades, which demonstrates that the diffusion postnet contributes to a better mel-spectrogram prediction.
We also visualize the generated mel-spectrograms. As shown in Figure \ref{fig:ablation1}, the diffusion-based postnet contributes to alleviating the mel-spectrogram over-smooth (see red box region).

\section{Conclusion}
In this paper, we propose RMSSinger, the first realistic-music-score-based singing voice synthesis (RMS-SVS) method, which utilizes the word-level modeling framework to avoid most tedious manual annotations. To achieve better pitch modeling, we propose the first diffusion-based pitch modeling method (P-DDPM), which incorporates the Gaussian diffusion and multinomial diffusion in a single model. Extensive experiments conducted on our collected dataset demonstrate the feasibility of our method for RMS-SVS and the superiority of our proposed P-DDPM.

\section{Acknowledgements}
This work was supported in part by the National Key R\&D Program of China under Grant No.2022ZD0162000,National Natural Science Foundation of China under Grant No.62222211, Grant No.61836002 and Grant No.62072397.

\section{Limitations}
There are majorly two limitations: Firstly, we collect a Chinese singing voice dataset and test our method only on this Chinese dataset due to the difficulty of recruiting professional singers in different languages. In the future, we will attempt to collect the singing voices dataset including more languages and test our method in multilingual settings. Secondly, our method adopts the diffusion model in pitch modeling and the postnet, which require multiple inference steps. We will try advanced acceleration methods for diffusion models in the future.

\section{Ethics Statement}
RMSSinger provides a high-quality realistic-music-score-based singing voice synthesis method, which may cause unemployment for people with related occupations. Furthermore, the possible misuse of realistic music scores from the website may lead to copyright issues. We will add some constraints to guarantee people who use our code or pre-trained model would not use the model in illegal cases.


\bibliography{my}

\begin{thebibliography}{34}
\expandafter\ifx\csname natexlab\endcsname\relax\def\natexlab#1{#1}\fi

\bibitem[{Dhariwal and Nichol(2021)}]{dhariwal2021diffusion}
Prafulla Dhariwal and Alexander Nichol. 2021.
\newblock Diffusion models beat gans on image synthesis.
\newblock \emph{Advances in Neural Information Processing Systems},
  34:8780--8794.

\bibitem[{Donahue et~al.(2020)Donahue, Dieleman, Bi{\'n}kowski, Elsen, and
  Simonyan}]{donahue2020end}
Jeff Donahue, Sander Dieleman, Miko{\l}aj Bi{\'n}kowski, Erich Elsen, and Karen
  Simonyan. 2020.
\newblock End-to-end adversarial text-to-speech.
\newblock \emph{arXiv preprint arXiv:2006.03575}.

\bibitem[{Gu et~al.(2021)Gu, Yin, Rao, Wan, Tang, Zhang, Chen, Wang, and
  Ma}]{gu2021bytesing}
Yu~Gu, Xiang Yin, Yonghui Rao, Yuan Wan, Benlai Tang, Yang Zhang, Jitong Chen,
  Yuxuan Wang, and Zejun Ma. 2021.
\newblock Bytesing: A chinese singing voice synthesis system using duration
  allocated encoder-decoder acoustic models and wavernn vocoders.
\newblock In \emph{2021 12th International Symposium on Chinese Spoken Language
  Processing (ISCSLP)}, pages 1--5. IEEE.

\bibitem[{Gulati et~al.(2020)Gulati, Qin, Chiu, Parmar, Zhang, Yu, Han, Wang,
  Zhang, Wu et~al.}]{gulati2020conformer}
Anmol Gulati, James Qin, Chung-Cheng Chiu, Niki Parmar, Yu~Zhang, Jiahui Yu,
  Wei Han, Shibo Wang, Zhengdong Zhang, Yonghui Wu, et~al. 2020.
\newblock Conformer: Convolution-augmented transformer for speech recognition.
\newblock \emph{arXiv preprint arXiv:2005.08100}.

\bibitem[{He et~al.(2022)He, Zhao, Ren, Liu, Huai, and Yuan}]{he2022flow}
Jinzheng He, Zhou Zhao, Yi~Ren, Jinglin Liu, Baoxing Huai, and Nicholas Yuan.
  2022.
\newblock Flow-based unconstrained lip to speech generation.

\bibitem[{Ho et~al.(2020)Ho, Jain, and Abbeel}]{ho2020denoising}
Jonathan Ho, Ajay Jain, and Pieter Abbeel. 2020.
\newblock Denoising diffusion probabilistic models.
\newblock \emph{Advances in Neural Information Processing Systems},
  33:6840--6851.

\bibitem[{Hoogeboom et~al.(2021)Hoogeboom, Nielsen, Jaini, Forr{\'e}, and
  Welling}]{hoogeboom2021argmax}
Emiel Hoogeboom, Didrik Nielsen, Priyank Jaini, Patrick Forr{\'e}, and Max
  Welling. 2021.
\newblock Argmax flows and multinomial diffusion: Learning categorical
  distributions.
\newblock \emph{Advances in Neural Information Processing Systems},
  34:12454--12465.

\bibitem[{Huang et~al.(2021)Huang, Chen, Ren, Liu, Cui, and
  Zhao}]{huang2021multi}
Rongjie Huang, Feiyang Chen, Yi~Ren, Jinglin Liu, Chenye Cui, and Zhou Zhao.
  2021.
\newblock Multi-singer: Fast multi-singer singing voice vocoder with a
  large-scale corpus.
\newblock In \emph{Proceedings of the 29th ACM International Conference on
  Multimedia}, pages 3945--3954.

\bibitem[{Huang et~al.(2022{\natexlab{a}})Huang, Cui, Chen, Ren, Liu, Zhao,
  Huai, and Wang}]{huang2022singgan}
Rongjie Huang, Chenye Cui, Feiyang Chen, Yi~Ren, Jinglin Liu, Zhou Zhao,
  Baoxing Huai, and Zhefeng Wang. 2022{\natexlab{a}}.
\newblock Singgan: Generative adversarial network for high-fidelity singing
  voice generation.
\newblock In \emph{Proceedings of the 30th ACM International Conference on
  Multimedia}, pages 2525--2535.

\bibitem[{Huang et~al.(2022{\natexlab{b}})Huang, Lam, Wang, Su, Yu, Ren, and
  Zhao}]{huang2022fastdiff}
Rongjie Huang, Max~WY Lam, Jun Wang, Dan Su, Dong Yu, Yi~Ren, and Zhou Zhao.
  2022{\natexlab{b}}.
\newblock Fastdiff: A fast conditional diffusion model for high-quality speech
  synthesis.
\newblock \emph{arXiv preprint arXiv:2204.09934}.

\bibitem[{Huang et~al.(2022{\natexlab{c}})Huang, Ren, Liu, Cui, and
  Zhao}]{huang2022generspeech}
Rongjie Huang, Yi~Ren, Jinglin Liu, Chenye Cui, and Zhou Zhao.
  2022{\natexlab{c}}.
\newblock Generspeech: Towards style transfer for generalizable out-of-domain
  text-to-speech synthesis.
\newblock \emph{arXiv preprint arXiv:2205.07211}.

\bibitem[{Jeong et~al.(2021)Jeong, Kim, Cheon, Choi, and Kim}]{jeong2021diff}
Myeonghun Jeong, Hyeongju Kim, Sung~Jun Cheon, Byoung~Jin Choi, and Nam~Soo
  Kim. 2021.
\newblock Diff-tts: A denoising diffusion model for text-to-speech.
\newblock \emph{arXiv preprint arXiv:2104.01409}.

\bibitem[{Kim et~al.(2021)Kim, Kong, and Son}]{kim2021conditional}
Jaehyeon Kim, Jungil Kong, and Juhee Son. 2021.
\newblock Conditional variational autoencoder with adversarial learning for
  end-to-end text-to-speech.
\newblock In \emph{International Conference on Machine Learning}, pages
  5530--5540. PMLR.

\bibitem[{Kong et~al.(2020)Kong, Kim, and Bae}]{kong2020hifi}
Jungil Kong, Jaehyeon Kim, and Jaekyoung Bae. 2020.
\newblock Hifi-gan: Generative adversarial networks for efficient and high
  fidelity speech synthesis.
\newblock \emph{Advances in Neural Information Processing Systems},
  33:17022--17033.

\bibitem[{Liu et~al.(2022)Liu, Li, Ren, Chen, and Zhao}]{liu2022diffsinger}
Jinglin Liu, Chengxi Li, Yi~Ren, Feiyang Chen, and Zhou Zhao. 2022.
\newblock Diffsinger: Singing voice synthesis via shallow diffusion mechanism.
\newblock In \emph{Proceedings of the AAAI Conference on Artificial
  Intelligence}, volume~36, pages 11020--11028.

\bibitem[{Lu et~al.(2020)Lu, Wu, Luan, Tan, and Zhou}]{lu2020xiaoicesing}
Peiling Lu, Jie Wu, Jian Luan, Xu~Tan, and Li~Zhou. 2020.
\newblock Xiaoicesing: A high-quality and integrated singing voice synthesis
  system.
\newblock \emph{arXiv preprint arXiv:2006.06261}.

\bibitem[{Miao et~al.(2020)Miao, Liang, Chen, Ma, Wang, and
  Xiao}]{miao2020flow}
Chenfeng Miao, Shuang Liang, Minchuan Chen, Jun Ma, Shaojun Wang, and Jing
  Xiao. 2020.
\newblock Flow-tts: A non-autoregressive network for text to speech based on
  flow.
\newblock In \emph{ICASSP 2020-2020 IEEE International Conference on Acoustics,
  Speech and Signal Processing (ICASSP)}, pages 7209--7213. IEEE.

\bibitem[{Nichol and Dhariwal(2021)}]{nichol2021improved}
Alexander~Quinn Nichol and Prafulla Dhariwal. 2021.
\newblock Improved denoising diffusion probabilistic models.
\newblock In \emph{International Conference on Machine Learning}, pages
  8162--8171. PMLR.

\bibitem[{Oord et~al.(2016)Oord, Dieleman, Zen, Simonyan, Vinyals, Graves,
  Kalchbrenner, Senior, and Kavukcuoglu}]{oord2016wavenet}
Aaron van~den Oord, Sander Dieleman, Heiga Zen, Karen Simonyan, Oriol Vinyals,
  Alex Graves, Nal Kalchbrenner, Andrew Senior, and Koray Kavukcuoglu. 2016.
\newblock Wavenet: A generative model for raw audio.
\newblock \emph{arXiv preprint arXiv:1609.03499}.

\bibitem[{Ren et~al.(2020{\natexlab{a}})Ren, Hu, Tan, Qin, Zhao, Zhao, and
  Liu}]{ren2020fastspeech}
Yi~Ren, Chenxu Hu, Xu~Tan, Tao Qin, Sheng Zhao, Zhou Zhao, and Tie-Yan Liu.
  2020{\natexlab{a}}.
\newblock Fastspeech 2: Fast and high-quality end-to-end text to speech.
\newblock \emph{arXiv preprint arXiv:2006.04558}.

\bibitem[{Ren et~al.(2021)Ren, Liu, and Zhao}]{ren2021portaspeech}
Yi~Ren, Jinglin Liu, and Zhou Zhao. 2021.
\newblock Portaspeech: Portable and high-quality generative text-to-speech.
\newblock \emph{Advances in Neural Information Processing Systems},
  34:13963--13974.

\bibitem[{Ren et~al.(2019)Ren, Ruan, Tan, Qin, Zhao, Zhao, and
  Liu}]{ren2019fastspeech}
Yi~Ren, Yangjun Ruan, Xu~Tan, Tao Qin, Sheng Zhao, Zhou Zhao, and Tie-Yan Liu.
  2019.
\newblock Fastspeech: Fast, robust and controllable text to speech.
\newblock \emph{Advances in Neural Information Processing Systems}, 32.

\bibitem[{Ren et~al.(2020{\natexlab{b}})Ren, Tan, Qin, Luan, Zhao, and
  Liu}]{ren2020deepsinger}
Yi~Ren, Xu~Tan, Tao Qin, Jian Luan, Zhou Zhao, and Tie-Yan Liu.
  2020{\natexlab{b}}.
\newblock Deepsinger: Singing voice synthesis with data mined from the web.
\newblock In \emph{Proceedings of the 26th ACM SIGKDD International Conference
  on Knowledge Discovery \& Data Mining}, pages 1979--1989.

\bibitem[{Ren et~al.(2022)Ren, Tan, Qin, Zhao, and Liu}]{ren2022revisiting}
Yi~Ren, Xu~Tan, Tao Qin, Zhou Zhao, and Tie-Yan Liu. 2022.
\newblock Revisiting over-smoothness in text to speech.
\newblock \emph{arXiv preprint arXiv:2202.13066}.

\bibitem[{Sohl-Dickstein et~al.(2015)Sohl-Dickstein, Weiss, Maheswaranathan,
  and Ganguli}]{sohl2015deep}
Jascha Sohl-Dickstein, Eric Weiss, Niru Maheswaranathan, and Surya Ganguli.
  2015.
\newblock Deep unsupervised learning using nonequilibrium thermodynamics.
\newblock In \emph{International Conference on Machine Learning}, pages
  2256--2265. PMLR.

\bibitem[{Song et~al.(2020)Song, Meng, and Ermon}]{song2020denoising}
Jiaming Song, Chenlin Meng, and Stefano Ermon. 2020.
\newblock Denoising diffusion implicit models.
\newblock \emph{arXiv preprint arXiv:2010.02502}.

\bibitem[{Umbert et~al.(2015)Umbert, Bonada, Goto, Nakano, and
  Sundberg}]{umbert2015expression}
Marti Umbert, Jordi Bonada, Masataka Goto, Tomoyasu Nakano, and Johan Sundberg.
  2015.
\newblock Expression control in singing voice synthesis: Features, approaches,
  evaluation, and challenges.
\newblock \emph{IEEE Signal Processing Magazine}, 32(6):55--73.

\bibitem[{Vaswani et~al.(2017)Vaswani, Shazeer, Parmar, Uszkoreit, Jones,
  Gomez, Kaiser, and Polosukhin}]{vaswani2017attention}
Ashish Vaswani, Noam Shazeer, Niki Parmar, Jakob Uszkoreit, Llion Jones,
  Aidan~N Gomez, {\L}ukasz Kaiser, and Illia Polosukhin. 2017.
\newblock Attention is all you need.
\newblock \emph{Advances in neural information processing systems}, 30.

\bibitem[{Wang et~al.(2018)Wang, Takaki, and
  Yamagishi}]{wang2018autoregressive}
Xin Wang, Shinji Takaki, and Junichi Yamagishi. 2018.
\newblock Autoregressive neural f0 model for statistical parametric speech
  synthesis.
\newblock \emph{IEEE/ACM Transactions on Audio, Speech, and Language
  Processing}, 26(8):1406--1419.

\bibitem[{Wang et~al.(2022)Wang, Wang, Zhu, Wu, Li, Xue, Zhang, Xie, and
  Bi}]{wang2022opencpop}
Yu~Wang, Xinsheng Wang, Pengcheng Zhu, Jie Wu, Hanzhao Li, Heyang Xue, Yongmao
  Zhang, Lei Xie, and Mengxiao Bi. 2022.
\newblock Opencpop: A high-quality open source chinese popular song corpus for
  singing voice synthesis.
\newblock \emph{arXiv preprint arXiv:2201.07429}.

\bibitem[{Wang et~al.(2017)Wang, Skerry-Ryan, Stanton, Wu, Weiss, Jaitly, Yang,
  Xiao, Chen, Bengio et~al.}]{wang2017tacotron}
Yuxuan Wang, RJ~Skerry-Ryan, Daisy Stanton, Yonghui Wu, Ron~J Weiss, Navdeep
  Jaitly, Zongheng Yang, Ying Xiao, Zhifeng Chen, Samy Bengio, et~al. 2017.
\newblock Tacotron: Towards end-to-end speech synthesis.
\newblock \emph{arXiv preprint arXiv:1703.10135}.

\bibitem[{Zhang et~al.()Zhang, Li, Wang, Deng, Liu, Ren, He, Huang, Zhu, Chen
  et~al.}]{zhangm4singer}
Lichao Zhang, Ruiqi Li, Shoutong Wang, Liqun Deng, Jinglin Liu, Yi~Ren,
  Jinzheng He, Rongjie Huang, Jieming Zhu, Xiao Chen, et~al.
\newblock M4singer: A multi-style, multi-singer and musical score provided
  mandarin singing corpus.
\newblock In \emph{Thirty-sixth Conference on Neural Information Processing
  Systems Datasets and Benchmarks Track}.

\bibitem[{Zhang et~al.(2022{\natexlab{a}})Zhang, Cong, Xue, Xie, Zhu, and
  Bi}]{zhang2022visinger}
Yongmao Zhang, Jian Cong, Heyang Xue, Lei Xie, Pengcheng Zhu, and Mengxiao Bi.
  2022{\natexlab{a}}.
\newblock Visinger: Variational inference with adversarial learning for
  end-to-end singing voice synthesis.
\newblock In \emph{ICASSP 2022-2022 IEEE International Conference on Acoustics,
  Speech and Signal Processing (ICASSP)}, pages 7237--7241. IEEE.

\bibitem[{Zhang et~al.(2022{\natexlab{b}})Zhang, Zheng, Li, and
  Lu}]{zhang2022wesinger}
Zewang Zhang, Yibin Zheng, Xinhui Li, and Li~Lu. 2022{\natexlab{b}}.
\newblock Wesinger: Data-augmented singing voice synthesis with auxiliary
  losses.
\newblock \emph{arXiv preprint arXiv:2203.10750}.

\end{thebibliography}
\bibliographystyle{acl_natbib}

\appendix

\section{More Experimental Details}
\label{sec:app_exp}
\subsection{Subjective Evaluation}
We randomly select 16 sentences from the test set for the subjective evaluation. Each ground-truth audio or generated audio has been listened to by at least 15 professional listeners. For MOS-P and CMOS-P, listeners are told to focus on the naturalness of pitch modeling (e.g., vibrato part, UV part, and so on). For MOS-Q and CMOS-Q, listeners are told to focus on audio quality (e.g., noise, high-frequency details, pronunciation, and so on). For MOS, each listener is asked to evaluate different audio samples on a 1 - 5 Likert scale. For CMOS, listeners are told
to compare pairs of audio generated by different systems and indicate which of the two audio they
prefer and following the rule: 0
indicating no difference, 1 indicating a small difference, and 2 indicating a large difference. 
 All listeners get equally paid.

\subsection{Training Details}
We train and evaluate our model on a single NVIDIA 2080Ti GPU. Adam optimizer is used with $\beta_1 = 0.9, \beta_2 = 0.98$. It takes 180000 steps for the first stage of training and 160000 steps for the second stage. It takes about 24 hours for each stage of training on a single NVIDIA 2080Ti GPU.

\section{More Model Details}
\label{sec:app_model}
\subsection{Encoder}
Our phoneme encoder consists of 1 phoneme embedding layer and 4 Feed-Forward Transformer (FFT) blocks. Each FFT block consists of 1 multi-head attention layer with 2 attention heads and 1 1D convolution layer with the kernel size set to 5. All hidden channels are set to 256. 

\subsection{Decoder}
Our mel decoder has a similar architecture to the phoneme encoder except that the mel decoder does not consist of the phoneme embedding layer.
\end{document}